\documentclass[11pt,a4paper]{article}
\pdfoutput=1

\usepackage{graphicx}
\usepackage{caption}
\usepackage{subcaption}
\usepackage{bm}
\usepackage{amsmath}
\usepackage{amssymb,xfrac}
\usepackage[bbgreekl]{mathbbol}
\usepackage{cite,color,float}

\begin{document}

\pagestyle{plain}

\begin{center}
~

\vspace{1cm} {\large \textbf{
First-Order Phase Transition by XY-Model of Particle Dynamics
}}

\vspace{1cm}

Amir H. Fatollahi

\vspace{.5cm}

{\it Department of Physics, Alzahra University, \\ P. O. Box 19938, Tehran 91167, Iran}

\vspace{.3cm}

\texttt{fath@alzahra.ac.ir}

\vskip .8 cm
\end{center}

\begin{abstract}
\noindent A gas-liquid type of phase transition is found 
based on the particle dynamics on radius-$R$ circle in which the coordinate appears 
as the angle-variable of 1D XY-model.
Due to the specific appearance of compact-space radius (volume) 
in the present interpretation of XY-model, the ground-state 
develops a minimum at some critical radius, leading to the \textit{multi-valued}
Gibbs energy similar to systems with first-order phase transition.
\end{abstract}

\vspace{1cm}



\newpage

\section{Introduction}
The absence of phase transition in one-dimensional models of magnetic systems 
is commonly considered as a special case of the so-called van~Hove's theorem \cite{vanhove}.
However, a detailed examination of conditions by the theorem shows that even 
1D models may exhibit phase transitions \cite{cuesta}.

In the present note a model is considered for particle dynamics 
based on the 1D XY-model of magnetic systems. The model was initially 
considered in \cite{spchfath}, with emphasize on the phase structure due to 
the model's defining-parameter. 
In the present work the main concern is the
first-order phase transition based on the $PV$ equation-of-state by the model, 
qualitatively similar to the gas-liquid phase transition for systems with
for example Van der Waals equation-of-state \cite{Huang}. 
In the proposed dynamics, the coordinates are assumed 
to be compact variables of radius-$R$ circles, appearing as angle-variables of 1D 
XY-models living on the particle's discrete worldline.
The present approach to particle dynamics is similar 
to the one in lattice formulation of gauge theories,
in which the gauge variables appear as compact angle-variables living on a discrete lattice.
As will be discussed in detail, due to the specific appearance of the radius 
in the formulation, the ground-state energy develops a 
minimum at some critical radius, leading to the multi-valued 
Gibbs energy quite similar to the systems with gas-liquid 
first-order phase transition. 

The organization of the rest of the paper is as follows. In Sec.~2
the model for particle dynamics as well as its exact spectrum and 
the emergence of minimum in the ground-state are presented.
In Sec.~3 the thermodynamics and the phase structure by 
the model are presented. It is discussed how the 
minimum in the ground-state leads to the $PV$ and $GP$-diagrams 
similar to those of gas-liquid systems. In Sec.~4 
a detailed comparison is made between the present and the magnetic 
interpretations of the XY-model. In particular, the distinguished role by 
the compact space radius in the announced first-order phase transition is clarified. 
Sec.~5 is devoted to concluding remarks and possible extensions of the present model. 

\section{The model and its spectrum}

In this section the model and its exact spectrum are presented. However,
it is useful to review the basic elements of the thermodynamics by 
the ordinary dynamics. It is known that the ordinary dynamics of free
particles does not lead to a phase transition. 
In particular, the one-particle partition function at temperature 
$T=\beta^{-1}$ for a particle of mass $m$ in a $d$-dimensional 
box of volume $V=L^d$ is given by  \cite{wipf}
\begin{align}\label{01}
Z_0(\beta,V )= 
\left(\frac{m }{2\pi \,a}\right)^{d/2}
\int_{-L/2}^{L/2}\prod_{i=1}^{d} \prod_{n=0}^{N-1}
d{x}^i_n~ e^{S_0}
\end{align}
in which $S_0$ is the imaginary-time action in the time-sliced form
\begin{align}\label{02}
S_0 = - \frac{m }{2\,a} \sum_{i=1}^d\sum_{n=0}^{N-1}(
x^i_{n+1}-x^i_n)^2 
\end{align}
with $a=\beta / N$ as the tiny time-slice parameter.
The representation (\ref{01}) is to be supplemented by the periodic 
condition $x^i_0=x^i_N$ (in the continuous-time form 
${x}^i(0)={x}^i(\beta)$). 
In the $L\to\infty$ limit (\ref{01}) reduces to the well-known expression \cite{Huang}
\begin{align}\label{03}
Z_0(\beta,V) = (2\pi m/\beta)^{d/2}\,V
\end{align}
By means of the free-energy $A=-T\ln Z$, the equation-of-state $P=-(\partial A/ \partial V)_T$ 
leads to $P=T/V$, by which one expects the thermodynamics of a 
\textit{single-phase} ideal gas. 

In the present work the particle dynamics based on the 1D XY-model 
is again introduced by the imaginary-time action, in 
which instead of the action (\ref{02}), we consider (in units $\hbar=c=1$)
\begin{align}\label{04}
S =\frac{m\,R^2}{a}
\sum_{i=1}^{d} \sum_{n}\left(\cos\frac{x^i_{n+1}-x^i_{n}}{R}-1\right)
\end{align}
in which ``$a$" appears as the spacing parameter on the discrete worldline.
The coordinate $x^i$ is treated as compact angle-variable for which we assume
\begin{align}\label{05}
-\pi R &\leq  x^i \leq \pi R
\end{align}
As mentioned earlier, the treatment of coordinates as angle-variables may be considered as the
continuation of the agenda originated by lattice formulation of gauge theories, in which 
the gauge fields act as angle variables \cite{lattice,kogut}. 
The action (\ref{04}) is in fact the sum of $d$ copies of 1D XY-model, for which it is known there is no phase transition as a magnetic system \cite{mattis}. 
As it will be clarified later, it is the specific appearance of the volume of 
compact space in the present interpretation, 
namely the presence of radius $R$ both inside and in front of 
the cosine functions in (\ref{04}),  which leads to the announced phase transition. 

The action (\ref{04}) reduces to the ordinary one (\ref{02}) in the limit 
$x^i/R\ll 1$. Using the transfer-matrix method one can obtain  
the energy spectrum by the imaginary-time (Euclidean) action. 
The element of transfer-matrix $\widehat{V}$ 
between two adjacent times $n$ and $n+1$ is given in terms of the imaginary-time
action \cite{wipf}
\begin{align}\label{06}
\langle  \vec{x}_{n+1} |\widehat{V} |  \vec{x}_n\rangle=\left(\frac{m }{2\pi a}\right)^{d/2}
\,\exp\left[\frac{mR^2}{a}
\sum_{i=1}^{d} \left(\cos\frac{x^i_{n+1}-x^i_{n}}{R}-1\right)\right]
\end{align}
In the Fourier basis $\langle \vec{x} | \vec{s} \rangle =
\exp(\mathrm{i}\,\vec{s}\cdot\vec{x}/R)/(2\pi R)^{d/2}$, 
it is easy to see that the above transfer-matrix is diagonal \cite{mattis,spchfath}. 
Using the relation $\widehat{V}=\exp(-a\widehat{H})$ with $\widehat{H}$ as
the Hamiltonian, one finds the exact energy spectrum \cite{mattis,spchfath}
\begin{align}\label{07}
E_{s_1,\cdots,s_d}(R)=-\frac{1}{a} 
\sum_{i=1}^d
\ln \left[\sqrt{\frac{2\pi mR^2}{a}  } \, e^{-mR^2/a}~I_{s_i}(mR^2/a )  \right]
\end{align}
in which $s_i$'s are integers, and $I_s$ is the modified Bessel function. 
In fact the expression for the spectrum of XY-model of magnetic systems is essentially as 
above \cite{mattis}, except for the extra square-root $\sqrt{2\pi mR^2/a}$, 
by which a key minimum is developed in the ground-state $E_{0,\cdots,0}$ 
at radius (see Fig.~1)
\begin{align}\label{09}
R_\star= 0.889 \sqrt{a/m}
\end{align}
The minimum in the ground-state is an indication that
the system exhibits a first-order phase transition, as we will see later. 

\begin{figure}[t]
	\begin{center}
		\includegraphics[scale=0.65]{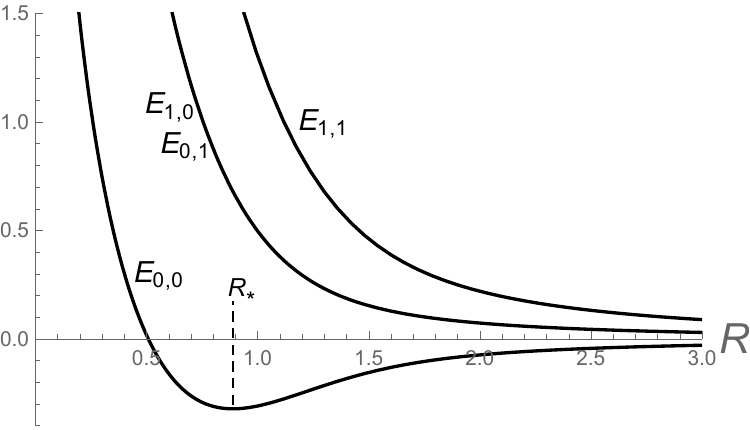}
	\end{center}
\vskip -.5cm
	\caption{\small The plots of three lowest energies in units $m=a=1$ for $d=2$ case.
The ground-state's minimum is at $R_\star\simeq 0.889$.}
\end{figure}

The limiting behaviors of the spectrum can be obtained. 
At the extreme large radius limit $mR^2/a\gg 1$, using 
$I_s(\alpha)\simeq \frac{e^{\alpha}}{\sqrt{2\pi \alpha}} 
\exp(-s^2/2\alpha)$ for $\alpha\gg s$, one finds the almost
continuous spectrum
\begin{align}\label{08}
E_{s_1,\cdots,s_d}\simeq\frac{1}{2mR^2}\,\sum_{i=1}^d s_i^2
\end{align}
as the ordinary kinetic energy of a free particle moving with 
momentum $p_i=s_i/R$ in the compact direction with radius $R$. 
In the small radius limit $mR^2/a\ll1$, using $I_s(\alpha)\simeq (\alpha/2)^2/s!$ for
$\alpha\ll1$, we find for the discrete spectrum
\begin{align}
E_{s_1,\cdots,s_d}\simeq \frac{1}{a}\ln\left(\frac{mR^2}{a}\right)
\sum_{i=1}^d (s_i+\frac{1}{2}) 
\end{align}
So the continuous spectrum in large radius limit approaches the 
discrete one in the small radius limit. 

\section{Partition function and phase transition}

The partition function may be evaluated either by the definition
\begin{align}\label{10}
Z (\beta,R):=\sum_{\{s_i\}} e^{-\beta\, E_{\{s_i\}}(R)}
\end{align}
or by means of the representation similar to (\ref{01}) \cite{wipf}
\begin{align}\label{11}
Z (\beta,R)&=\mathrm{Tr} \,\hat{V}^\beta \nonumber \\
&=\left(\frac{m }{2\pi a}\right)^{d/2}
\int_{-\pi R}^{\pi R}\, \prod_{i,\, n} d x^i_n 
\,\exp\left[\frac{mR^2}{a}
\sum_{i=1}^{d} \sum_{n=0}^{\beta-1}\left(\cos\frac{x^i_{n+1}-x^i_{n}}{R}-1\right)\right]
\end{align}
supplemented by the periodic condition $x_0=x_\beta$.
In the present case the equivalence of (\ref{10}) and (\ref{11}) is 
checked numerically. By either (\ref{10}) or (\ref{11}) it is easy to see that
\begin{align}\label{12}
Z =\left(Z_{d=1}\right)^d
\end{align}
simply because the action (\ref{04}) is fully separable in $x^i$'s.
By means of the free-energy $A =-T\ln Z $, the partition function (\ref{10}) can be 
used to study the phase structure by the model.
Hereafter we work with the choice $m=a=1$, and consider the case $d=3$, 
in which the volume and radius are related as 
$V=(2\pi\,R)^3$ or $R=V^{1/3}/2\pi$.
\begin{figure}[t]
	\begin{center}
		\includegraphics[scale=0.5]{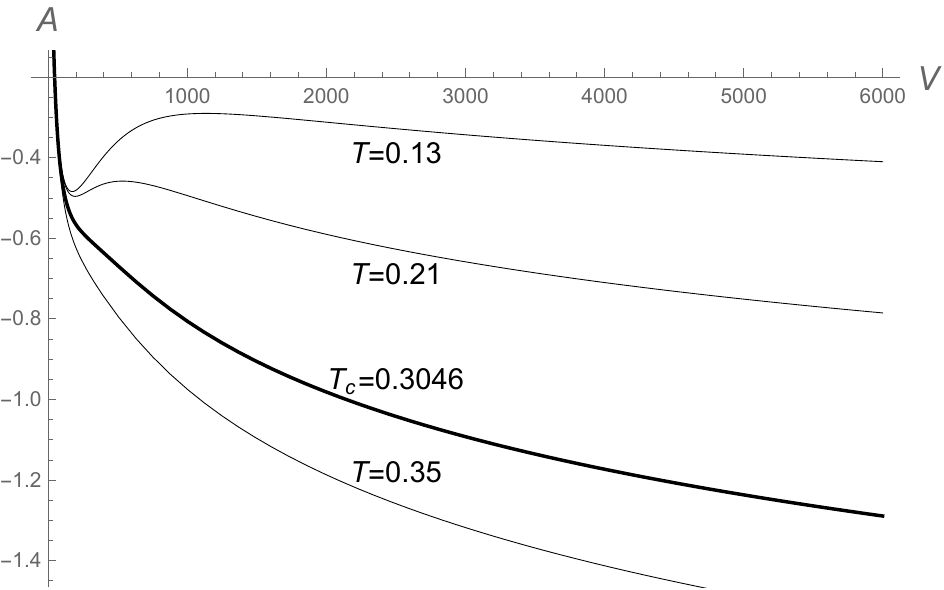}
	\end{center}
\vskip -.5cm
	\caption{\small The isothermal $AV$-diagrams for $d=3$ (temperature units $a^{-1}$). 
The diagrams with $T<T_c=0.3046$ consist a part in which there are 
points with common tangents (slopes), leading to equal pressures.}
\end{figure}
As the consequence of the mentioned minimum in the ground-state
at radius $R_\star$, at sufficiently low temperatures where the 
ground-state has considerable contribution to $Z$, 
there are points with common tangents (slopes) in the isothermal $AV$-diagrams.
In Fig.~2 the isothermal $AV$-diagrams clearly confirm this expectation
below the critical temperature $T_c=0.3046$. 
With pressure as slope by $P=-(\partial A/ \partial V)_T$, 
in the isothermal $PV$-diagrams there should be different volumes with equal 
pressure, leading to parts with positive slope $\partial P/\partial V>0$. 
The positive slope in $PV$-diagram would mean a negative 
compressibility, leading to the mechanical instability of the system \cite{Huang}.
The mentioned behaviors are similar to those by the 
van~der~Waals equation-of-state, commonly used to 
describe the systems with gas-liquid transition \cite{Huang,stanley}. 
In the real situation, the system has a constant pressure during the gas-liquid 
transition, the so-called vapor-pressure \cite{Huang,stanley}. 
To reflect the real situation, the $PV$-diagrams are to be modified 
based on the so-called Maxwell construction \cite{Huang,stanley},
by which the constant pressure part finds a thermodynamical basis. 
To maintain the stability of system at equilibrium, 
a proper treatment of the part with positive $\partial P/\partial V$ is essential.

\begin{figure}[t]
	\begin{center}
		\includegraphics[scale=0.4]{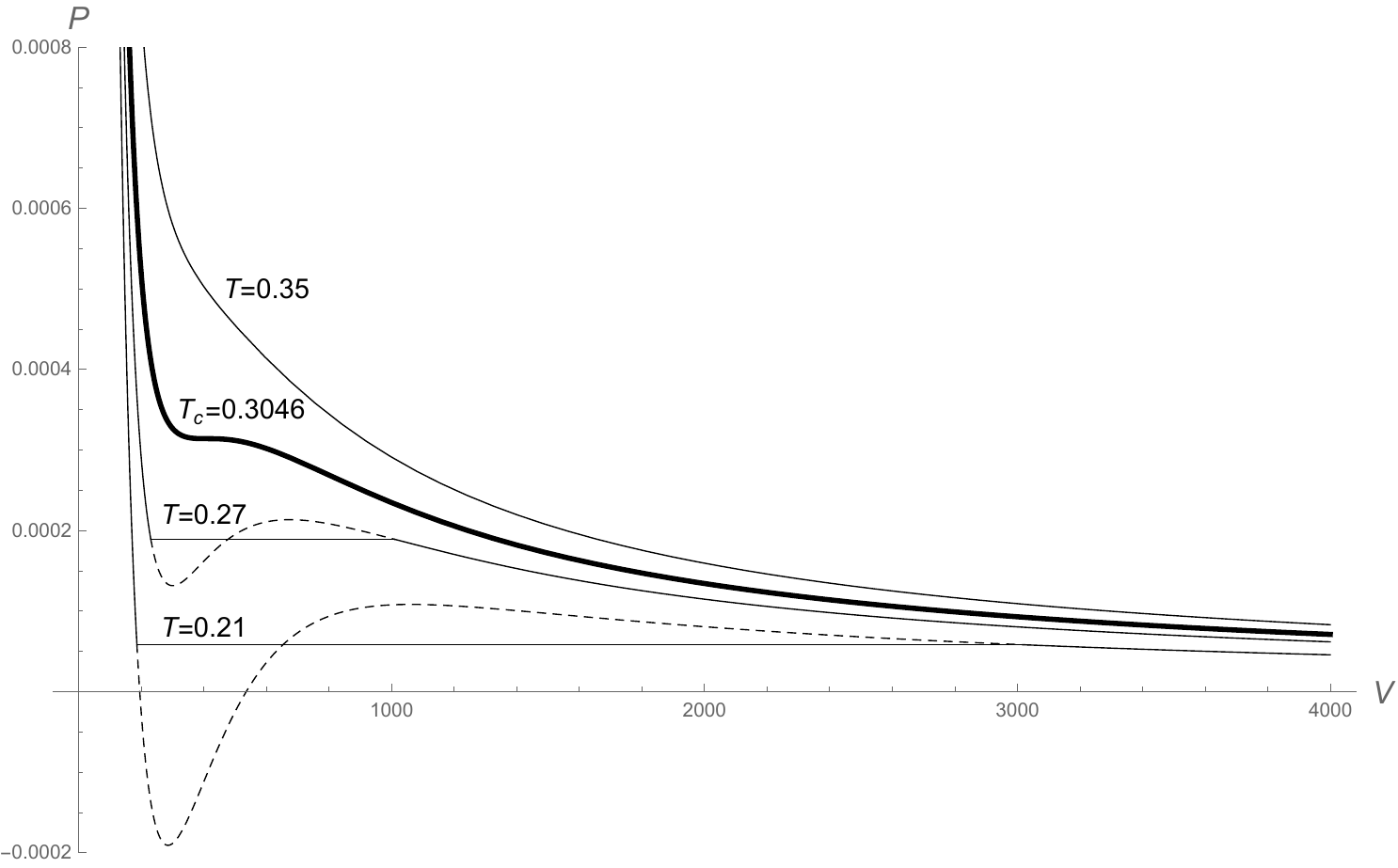}
	\end{center}
\vskip -.5cm
	\caption{\small The isothermal $PV$-diagrams at different temperatures.
The dashed parts represent the diagrams before the modification by the Maxwell construction. }
\end{figure}

%
\begin{figure}[t]
	\begin{center}
		\includegraphics[scale=0.5]{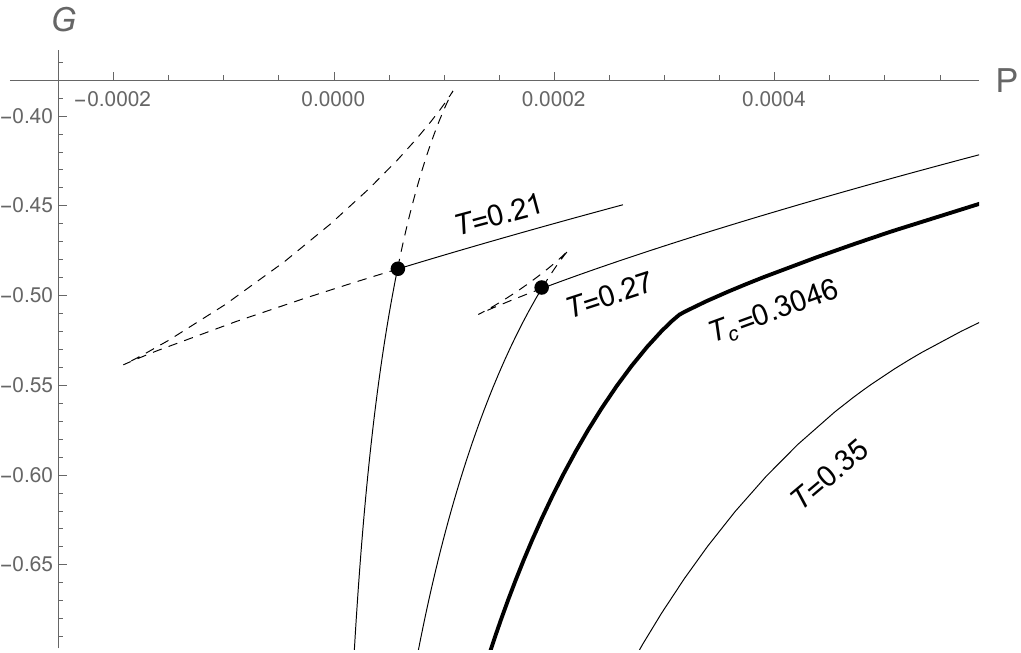}
	\end{center}
\vskip -.5cm
	\caption{\small The isothermal $GP$-diagrams. The dashed lines represent the 
parts that are not being followed by system, leading to jump in slope at cusp.}
\end{figure}

In Fig.~3 samples of $PV$-diagrams by the present model are plotted, in which both 
unmodified paths (dashed lines) and modified ones are present. 
The Maxwell construction and the corresponding phase structure 
are best described based on the Gibbs energy $G=A+PV$. 
As the consequence of the mentioned behavior of free-energy, 
the isothermal $GP$-diagrams develop cusps below $T_c$ 
\cite{Huang, stanley}, after which $G$ is \textit{multi-valued} for some pressures. 
For the present model 
plots of the isothermal $GP$-diagrams are presented in Fig.~4, 
in which the expected cusps are evident.
As for states with equal temperature and pressure, the state with lower $G$ 
is selected by the system \cite{Huang}, the parts beyond the cusps are not followed by 
the system. Instead, as volume is changed the constant pressure at cusp 
is hold during a phase transition. 
In Fig.~4 the dots at cusps indicate the constant vapor-pressure values, and 
each dashed part in Fig.~3 corresponds to a similar part in Fig.~4, 
both not being followed by the system.

\begin{figure}[t]
	\begin{center}
		\includegraphics[scale=0.5]{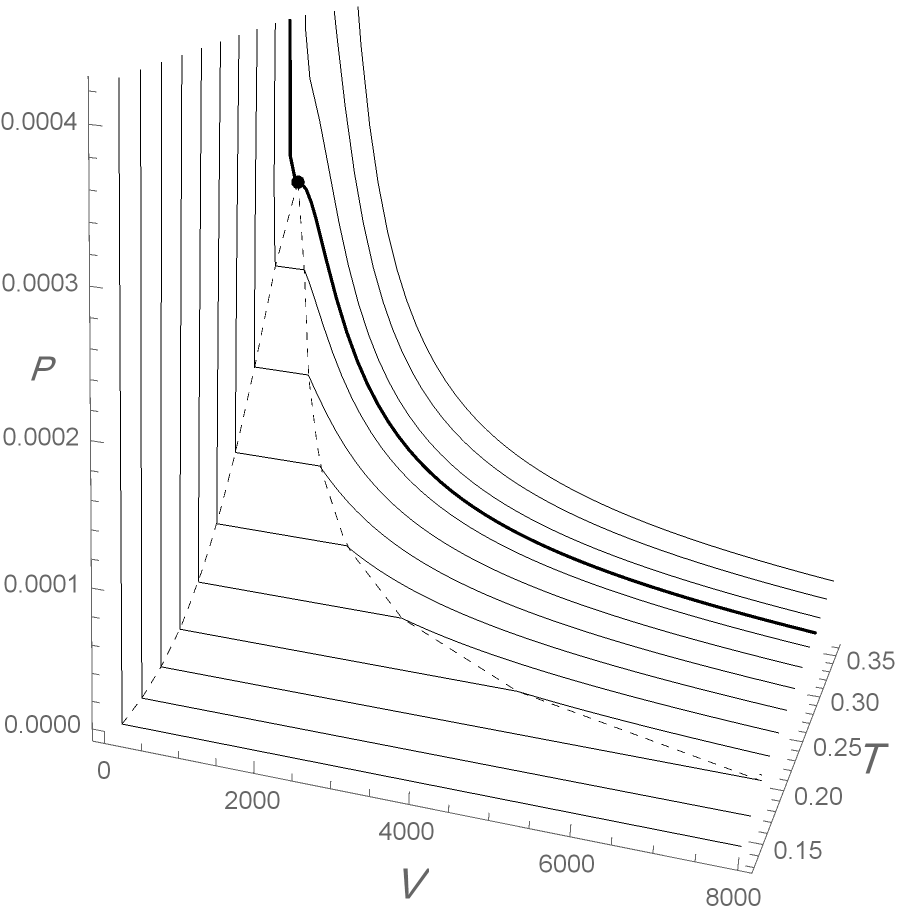}
	\end{center}
\vskip -1cm
	\caption{\small The $PVT$-surface of states for $d=3$. The thick curve is at critical temperature
$T_c=0.3046$, with the critical point (black dot) coordinates $P_c=0.000314$ and $V_c=409.6$.}
\end{figure}

As at the cusp there is a jump in the first derivative $\partial G / \partial P$,
the corresponding transition is categorized as a first-order phase transition
\cite{Huang, stanley}. The $PVT$-surface for the present model 
is presented in Fig.~5, which is quite reminiscent to the one of 
a system with the gas-liquid phase transition \cite{Huang}.

\section{The origin of the minimum in ground-state}

The 1D XY-model of magnetic systems is well known to exhibit no phase 
transition. As in the present interpretation 
the system is reduced to $d$ copies of 1D model, it is necessary 
to understand the origin of the phase transition by the present interpretation of the model. 
The 1D XY-model of magnetic systems is given by the Hamiltonian
\begin{align}\label{13}
H_\mathrm{Mag}= -J\sum_{n}\big(\cos(\theta_{n+1}-\theta_n)-1\big)
\end{align}
with $J$ as the coupling constant, and $\theta_n$ as the orientation 
of $n$-th classical spin on the 1D chain. 
The partition function for the magnetic system at temperature 
$T=1/\beta$ is then given by \cite{mattis}
\begin{align}\label{14}
Z_\mathrm{Mag} = \int_{-\pi}^{\pi}\, \prod_{n} d\theta_n 
\,\exp\!\Big[\beta\,J
\sum_{n}\big(\cos(\theta_{n+1}-\theta_{n})-1\big)\Big]
\end{align}
Now the key point is, in the magnetic interpretation of the model 
the coupling $J$ is absent inside 
the cosine functions of (\ref{13}) and (\ref{14}), in contrast to the situation in 
the particle dynamics interpretation, in which the radius $R$ appears both in front 
and inside the cosine functions
of (\ref{04}) and (\ref{11}). The way of appearance of radius $R$ makes the 
fundamental difference between two interpretations of the 1D XY-model. 
In the particle dynamics interpretation one may get rid of $R$ inside the cosine function
by defining the angle variables $\theta_n=x_n/R$ with 
$-\pi\leq \theta_n \leq \pi$. However, this does not remove 
$R$'s completely, as they would rise in the integral-measure 
of (\ref{11}), explicitly as 
\begin{align}\label{15}
Z_{d=1}=
\left(\frac{m }{2\pi a}\right)^{d/2}
\int_{-\pi}^{\pi}\, \prod_{n} \big(R~ d\theta_n\big) 
\,\exp\left[\frac{mR^2}{a}
 \sum_{n=0}^{\beta-1}\big(\cos(\theta_{n+1}-\theta_{n})-1\big)\right]
\end{align}
by which the different appearance of $R$ in comparison with $J$ in (\ref{14}) is observed.
Concerning the energy spectrum, as mentioned earlier it is easy to check that the presence of 
$\sqrt{mR^2/a}$ in energy (\ref{07}) is responsible for the minimum in the ground-state.
This extra appearance of $R$ does not have an analog for $J$ 
in the spectrum of 1D magnetic system \cite{mattis}. Now again, 
by the new variables $\theta_n=x_n/R$ with the Fourier basis 
$\langle \theta | s \rangle =\exp(\mathrm{i}\,s\,\theta)/\sqrt{2\pi}$,
using the relation $|x\rangle = |\theta\rangle/\sqrt{R}$ the radius $R$ comes back 
as a pre-factor for transfer-matrix (\ref{06}).
In summary, in the particle dynamics interpretation of XY-model, 
in contrast to the 1D magnetic system with no phase transition, 
the specific appearance of compact space radius (volume) 
in the model is the origin of the phase transition. 

\section{Conclusion and Discussion}

In this note a particle dynamics interpretation of the XY-model is considered,
in which the coordinates are assumed to be compact variables, appearing
as angle-variables of the 1D XY-model. The present interpretation may be
considered as a continuation of the agenda leading to the 
lattice formulation of gauge theories, in which the gauge fields 
are assumed to be compact angle-variables \cite{lattice,kogut}. Accordingly, this leads to 
the replacement of the action of the quadratic form 
$\frac{1}{4}\int \! F_{\mu\nu}^2$ by the cosine form 
$\frac{1}{2}\sum_P (1-\cos \Phi_P)$, with $\Phi_P$ as the
field-flux inside the plaquette $P$. 
In the same way, here the quadratic form (\ref{02}) for action of 
a particle is replaced with (\ref{04}), leading to the mentioned phase structure. 

In the present interpretation it is clarified that the specific treatment of‎ 
‎the dynamics on compact space in the 1D XY-model leads to a first order phase transition‎, 
‎similar to the $PV$-diagrams of gas-liquid systems‎. 
In fact there are examples originated from the 2D magnetic systems that
may exhibit a phase transition of the first-order. In \cite{domany,blote} a power-form
of the interaction-terms in 2D XY and Heisenberg models are 
considered, leading to a first-order phase transition for sufficiently large power being used. 
In the present case, however, the first-order phase transition is exhibited 
by the same power-one cosine-term of 1D XY-model due to the different 
interpretation. 

Apart from the theoretical aspects and implications, 
one may try to find some applications for the present interpretation, 
for which the above-mentioned relation with lattice gauge theory may 
come useful. It is known that the lattice formulation of gauge theories
approach the ordinary formulation at weak coupling limit $g\ll1$, 
and the behavior at strong coupling limit is expected to be described 
by the lattice formulation. In the presented particle dynamics based 
on the angle-variables also the ordinary formulation is recovered at large 
radius limit $R\to \infty$ by (\ref{09}), and the possible relevance of
the formulation would appear at the small-$R$ limit. This leads to a possible 
application of the present model in the cosmological context. In the
present cosmological models it is assumed that the matter ingredients 
obey the ordinary dynamics with the known thermodynamics. It
is of interest to see how the dynamics and phase structure by the present model 
affects the different stages of Universe during the expansion, through 
which the system is driven from extremely small sizes to its present 
extent. 

The extensions of the present model to magnetic systems other 
than the XY-model, such as Potts model with $\mathbb{Z}_n$-valued
spins, can be of interest. In the present work the system obeyed the 
Maxwell-Boltzmann statistics. The extensions to Fermi-Dirac and Bose-Einstein 
statistics may lead to interesting features. 

\vspace{3mm}
\textbf{Acknowledgement}: 
This work is supported by the Research Council of Alzahra University.



\begin{thebibliography}{99}

\bibitem{vanhove} L. van Hove, Physica \textbf{16} (1950) 137.

\bibitem{cuesta} J.A. Cuesta and A. Sanchez, 
J. Stat. Phys. \textbf{115} (2004) 869. 

\bibitem{spchfath} A.H. Fatollahi, 
Eur. Phys. J. C \textbf{77} (2017) 159.

\bibitem{Huang} K. Huang, ``Statistical Mechanics", Wiley 1987.

\bibitem{wipf} A. Wipf, ``Statistical Approach to Quantum Field Theory", 
Springer 2013, Sec.~8.5.1.

\bibitem{lattice} K.G. Wilson, 
Phys. Rev. D \textbf{10} (1974) 2445.

\bibitem{kogut} J.B. Kogut, 
Rev. Mod. Phys. \textbf{51} (1979) 659.

\bibitem{mattis} D.C. Mattis, 
Phys. Lett. A \textbf{104} (1984) 357.


\bibitem{stanley} H.E. Stanley, ``Introduction to Phase Transitions and Critical Phenomena",
Oxford Univ. Press 1971, Sec.~2.5.

\bibitem{domany}
E. Domany, M. Schick, and R.H. Swendsen, Phys. Rev. Lett. \textbf{52} (1984) 
 1535.

\bibitem{blote}
H.W.J. Blote, W. Guo, and H.J. Hilhorst, Phys. Rev. Lett. \textbf{88} (2002) 047203.


\end{thebibliography}
\end{document}